\documentclass[amsmath,amssymb,aps,pre,reprint, showpacs, superscriptaddress]{revtex4-1}

\usepackage{amsmath,amsfonts,amssymb,amsthm,bm}
\usepackage{enumerate}
\usepackage[inline]{enumitem}
\usepackage{graphicx}
\usepackage{hyperref}
\usepackage{mathtools}
\usepackage[T1]{fontenc} \usepackage{lmodern}
\usepackage{tikz}
\usepackage{tikz-3dplot}
\usepackage{tkz-euclide}

\usepackage{xcolor}

\usepackage{changes}

\begin{document}

\title[]{\textcolor{black}{Information conduction and convection in noiseless Vicsek flocks}}

\author{Daniel Gei\ss} \affiliation{Institute for Theoretical Physics, University of Leipzig, Germany} 
\affiliation{Max Planck Institute for Mathematics in the Sciences, Leipzig, Germany}
\author{Klaus Kroy}  \affiliation{Institute for Theoretical Physics, University of Leipzig, Germany}
\author{Viktor Holubec} \affiliation{Faculty of Mathematics and Physics, Charles University, Praha, Czech Republic}
\email{viktor.holubec@mff.cuni.cz}

\begin{abstract}
Physical interactions generally respect certain symmetries, such as reciprocity and energy conservation, which survive in coarse grained isothermal descriptions. Active many-body systems usually break such symmetries intrinsically, on the particle level, so that their collective behavior is often more naturally interpreted as a result of information exchange. Here, we study numerically how information  spreads from a ``leader'' particle through an initially aligned flock, described by the Vicsek model \textcolor{black}{without noise}. In the low-speed limit of a static spin lattice, we find purely conductive spreading, reminiscent of heat transfer. Swarm motility and heterogeneity can break reciprocity and spin conservation. But what seems more consequential for the swarm response is that the dispersion relation acquires a significant convective contribution along the leader's direction of motion.  
\end{abstract}


\maketitle


\section{Introduction}

Transfer of information, energy, or mass through complex interacting networks is of ubiquitous interest in many scientific disciplines. As examples think of the www and social media~\cite{lerman2010information,moreno2004dynamics}, epidemics~\cite{pastor2001epidemic,boguna2002epidemic,hethcote2000mathematics,levis2020flocking}, or heat conduction and diffusion~\cite{a1993heat,joseph1989heat,brodkey2003transport}.  
In particular, information, rather than the elementary physical interactions transmitting it, is key to groups of motile living agents, such as bird flocks~\cite{cavagna2018physics,procaccini2011propagating} or bacterial colonies~\cite{zhang2010collective,ben1994generic}. 
To understand the behavior of such far-from-equilibrium many-body systems is a main task of the surging field of active matter~\cite{gompper2020,bechinger2016active,ramaswamy2010mechanics}. Many new interesting phenomena have been uncovered, including motility-induced phase separation (MIPS)~\cite{cates2015motility}, and related forms of self-organization~\cite{hagan2016emergent,bauerle2018self} and 
pattern formation~\cite{farrell2012pattern,solon2015pattern}. Such studies could eventually lead to the development of novel types of ``smart (meta-)materials''~\cite{brambilla2013swarm,vernerey2019biological}. Yet, systematic studies of the mechanisms of information spreading through active matter systems are still scarce. 

In this work, we therefore analyze the information spreading in a two-dimensional Vicsek model (VM)~\cite{vicsek1995novel}, which is a paradigmatic model of dry active matter (without momentum conservation in the solvent)~\cite{ginelli2016physics,vicsek2012collective}. It provides a minimalistic description of active collective phenomena such as the formation of bird flocks or insect swarms. The VM resembles a magnet consisting of $N$ spins, which describe the orientations of the self-propelled particles. Their positions advance at constant speed, while their orientations are subject to mutual alignment interactions with their neighbors. Compared to the limit of interacting lattice spins or also to the case of digital information transport through disordered static networks (frequently studied in network theory) the VM is capable of more complex behavior. Its neighbor configurations are neither regular nor static but constitute a dynamical graph~\cite{kuhn2011dynamic,clementi2015information}.  As a consequence, information in the VM spreads not only by conduction but also by convection, hitchhiking with the motile particles~\cite{levis2020flocking}. Moreover, the information about particle positions and orientations is continuous, not digital.

In the following, we try to disentangle the various complications, by first studying information spreading on a static square lattice. For vanishing noise, this limit allows for an exact solution, which simplifies the analysis and provides good insight. Then we investigate the full deterministic (no noise) VM with nonzero velocity. For both cases, we study the information spreading for a scenario known as \textit{flooding}  in network theory~\cite{clementi2009information,clementi2015information,van2002flooding,melnik2002similarity}:  starting in an orientationally ordered state with a single ``leader'' particle that deviates from the rest, we investigate how its perturbing effect spreads to the others.  So far, flooding dynamics was mostly studied for static graphs; but see Ref.~\cite{clementi2015information} for a more general approach. To assess the spatio-temporal information spreading in the VM, we numerically determine the corresponding dispersion relation. Naturally, the convective flooding due to particle motion is found to dominate over conduction at higher speeds and over long distances. But it also gives rise to a considerable forward-backward symmetry breaking, rendering the dispersion relation spatially highly non-isotropic.

The paper is structured as follows: In Sec.~\ref{sec:VM}, we introduce the VM. The  zero-speed limit of the VM is discussed in Sec.~\ref{Sec: Lattice model}, which introduces the two flooding scenarios considered in this work: the \emph{firm leader} with constrained spin orientation, which eventually guides the flock into a new direction; and the \emph{lax leader} that delivers an initial impulse but afterwards relaxes freely, like all other spins. Finally, in Sec.~\ref{sec:finite_velocity}, we consider the general case of non-vanishing particle speeds, where the dispersion relation becomes  ambiguous, before we conclude in Sec.~\ref{sec:Conclusion}.

\section{Vicsek model} \label{sec:VM}

Since its introduction in 1995, many modifications of the original VM have been discussed in the literature~\cite{chate2008modeling}. Here, we consider the deterministic discrete-time variant describing $N$ particles self-propelling with  constant speed $v_0$ in 2 dimensions with topological alignment interactions. The position $\mathbf{r}_i(t)$ and velocity \textcolor{black}{$\mathbf{v}_i(t)$} of $i$th particle obey the dynamical equations 
\begin{align}
	\mathbf{v}_i(t+1) &= v_0 \Theta\Bigl[
	\mathbf{v}_i(t) + \mbox{$\sum_{j\neq i }$} n_{ij}(t) \mathbf{v}_j(t) 
	\Bigr],
	\label{eq:vtdisc}\\
	\mathbf{r}_i(t+1) &= \mathbf{r}_i(t) + \mathbf{v}_i(t+1),
	\label{eq:rtdisc}
\end{align}
where $\Theta(\mathbf{v}) \equiv \mathbf{v}/|\mathbf{v}|$ normalizes the velocity. The connectivity matrix $n_{ij}(t)$ defines the interaction network. We assume topological interactions: each particle interacts with its $N_\mathrm{int}$ nearest neighbors at time $t$. For these $n_{ij}(t)=1$, while it  vanishes otherwise. We have tested that  metric interactions, where each particle interacts with all neighbors within a given spatial distance, leads to qualitatively the same results (data not shown). In contrast to the standard VM, we neglect the noise.

\textcolor{black}{
Instead of using the particle velocities $\mathbf{v}_i(t)$ to characterize the system state, one can equivalently describe it by the angular variables $\theta_i(t)$, defined by $\mathbf{v}_i(t) = v_0 (\cos \theta_i, \sin \theta_i)$. In this language, Eq.~\eqref{eq:vtdisc} assumes the form~\cite{degond2014hydrodynamics}
\begin{equation}
    \theta_i(t+1) = \theta_i(t) + \frac{1}{N_i(t)} \mbox{$\sum_{j\neq i }$} n_{ij}(t) \sin\left(\theta_j(t)-\theta_i(t)\right), \label{eq:VMangle}
\end{equation}
where $N_i(t) \equiv v_0^{-1} | \mathbf{v}_i(t) + \mbox{$\sum_{j\neq i }$} n_{ij}(t) \mathbf{v}_j(t) | $ stems from the normalization in Eq.~\eqref{eq:vtdisc}.}

We consider the situation where one of the particles (the leader) in a completely polarized system suddenly changes its direction and initiates a collective maneuver~\cite{attanasi2014information,cavagna2015flocking,cavagna2015silent,cavagna2018physics}, due to the spreading of information about its flight direction through the flock~\cite{heppner1997three,treherne1981group}. To analyze the spreading of information in different directions with respect to the leader's velocity, it is useful to position it initially, at time $t=0$, in the center of the flock.  In the next section, we investigate the information transfer for the static spin-lattice ($v_0=0$), where the information only spreads by conduction. The interplay of conduction and convection, appearing for nonzero particle velocity, is then addressed in Sec.~\ref{sec:finite_velocity}.

\section{Zero-velocity limit of the VM}
\label{Sec: Lattice model}

\subsection{\textcolor{black}{Linearized} lattice VM and its continuous limits}

To make contact with classical spin models,
the particles are placed on grid points $\mathbf{r}_i$ of a two dimensional square lattice and interact only with their direct neighbors. \textcolor{black}{In this limit, the dynamics of the well-known XY model is restored. In the following, we label the orientations  $\theta_k$ of the individual spins (or particles) by their positions $ij$ in the lattice. If all spins on the square lattice are well aligned, Eq.~\eqref{eq:VMangle} can be expanded in fluctuations around the aligned state as} 
\begin{multline}
	\theta_{ij}(t+1) = \frac{1}{5} \left[ \theta_{ij}(t) + \theta_{i-1j}(t) + \theta_{ij-1}(t) \right. \\ 
	\left.
	+ \theta_{i+1j}(t) + \theta_{ij+1}(t) \right],
	\label{eq: average}
\end{multline}
\textcolor{black}{
where we assumed that the average orientation of the system is 0 and $\theta_{ij} \ll 1$. In this limit, the periodic boundary conditions for $\theta_{ij}$ do not need to be taken into account.
An analogous linear formulation of the low-velocity VM has recently been employed~\cite{de2022inferring} to calculate the total number of particles in a Vicsek flock from the orientational diffusion coefficient of a single particle.}

Noteworthy, the same equation describes occupation probabilities of the individual grid-points for a symmetric random walk on a two dimensional square lattice with equal probabilities to stay at a given point or to jump to a neighboring point. Unlike the standard Vicsek model, it thus conserves the total amount of `information' $\sum_{ij} \theta(t)$ unless some of the lattice points serve as sources or sinks of information. Information conservation would also be lost for less symmetric lattices, breaking reciprocity of the interactions (for details, see App.~\ref{sec:info_conserv}).

Besides being exactly solvable, the importance of this simplified lattice model for understanding of information transfer in the VM is its similarity to other physical models such as lattice models of ferromagnetism, where $\theta_{ij}(t)$ describes spin of the given grid point~\cite{joyce1967classical,kosterlitz1974critical,sentef2007spin,lima2009dynamics,sonin2010spin}, Google Search PageRank-Algorithm~\cite{page1999pagerank, avrachenkov2006effect}, measuring the importance of a web page by counting all links to it and weighting them by their quality, the majority vote model, and, most importantly, lattice models of heat conduction~\cite{sobolev1991transport,sobolev2017discrete}. 

A central finding from the latter is that the heat flux is well described by Fourier's law
implying that the local temperature $\theta$ obeys the parabolic (diffusion) 
\begin{equation}
	\partial_t \theta = D \nabla^2 \theta
	\label{eq: parabolic}
\end{equation}
with the diffusion coefficient $D$. However, this equation leads to unphysical infinite propagation speed of heat~\cite{sobolev1991transport,lebon2008understanding}, in the sense that a change in the temperature at the origin leads to infinitesimal changes in temperature far away from the origin after an infinitesimally short time. Another issue is that Eq.~\eqref{eq: parabolic} in
general cannot describe the propagation of second sound, i.e. the thermal wave~\cite{joseph1989heat} encountered in low-temperature physics~\cite{landau1941theory}. The most popular and simplest generalization of Eq.~\eqref{eq: parabolic} which can describe both diffusive and wave-like transfer is the hyperbolic equation
\begin{equation}
\partial_t \theta + \frac{\tau}{2} \partial_t^2 \theta  = \frac{\tau}{2} c^2 \nabla^2 \theta,
	\label{eq: hyperbolic}
\end{equation}
with maximum heat transfer velocity $c$ and a characteristic time $\tau$. A standard derivation of this equation is based on Cattaneo's generalization of Fourier's law~\cite{goldstein1951diffusion, cattaneo1958form}.

Interestingly, it turns out that both these equations are special limiting case of Eq.~\eqref{eq: average}~\cite{sobolev1991transport,sobolev2017discrete}. Specifically, introducing a lattice constant $\ell$ and the time $\tau$ the signal needs to travel between two lattice points, it can be rewritten as
	\begin{multline}
	\theta(x,y,t+\tau) = \frac{1}{5} \left[ \theta(x,y,t) + \theta(x+\ell,y,t) +\right.\\ \theta(x-\ell,y,t) +  \theta(x,y+\ell,t) + \left. \theta(x,y-\ell,t) \right] .
	\label{eq: continuous}
	\end{multline}
Now, taking the continuum limit $\tau \to 0$ and $\ell \to 0$, while keeping constant the ratio $5 D \equiv\ell^2/\tau$ yields in the zeroth order in $\tau$ the diffusion equation~\eqref{eq: parabolic}. On the other hand, taking the limit while keeping constant the velocity $c\sqrt{5/2}\equiv\ell/\tau$ leads in the first order in $\tau$ to the hyperbolic equation~\eqref{eq: hyperbolic}. These non-standard definitions of speed and diffusion coefficient result from the term $\theta(x,y,t)$ on the right-hand side of Eq.~\eqref{eq: continuous}, which is not present in a standard random walk. 
\textcolor{black}{The} speed $c$, denoting the maximum speed of propagation in Eq.~\eqref{eq: hyperbolic}, is smaller than the maximum speed of propagation in the lattice model, $v=l/\tau$. Identifying $\tau c^2/2$ in Eq.~\eqref{eq: hyperbolic} with $D$, one can consider the parabolic equation~\eqref{eq: parabolic} as a limit of infinitely fast ($\tau = 0$) signal transmission between the neighboring lattice points. While this limit is often a good approximation for heat conduction~\cite{joseph1989heat} it might not be appropriate for biological agents with finite response time. 
\textcolor{black}{A general statement about which of the two continuum limits fits better the description of the VM is not possible. It can heavily depend on the quantity of interest and the chosen parameters. Nonetheless, from our analysis below, it follows that the information spreading in the VM is approximately diffusive for small speeds $v_0$ and increasingly non-diffusive as $v_0$ grows.}
 
\subsection{Firm and lax leaders} \label{Sec: LM Numerical}

We now consider the following two specific flooding scenarios  for the static VM~\eqref{eq: average}. (a) In the firm leader scenario,  \textcolor{black}{the leader's orientation is held fixed. Measuring the angular variables $\theta_{ij}\ll 1$ in units of the initial orientation of the leader, we set $\theta_{00}(t)=1$ for all times.} This amounts to a steady information influx into the system. (b) In the lax leader scenario, the orientation of the leader is set to 1 at time 0 but then evolves according to Eq.~\eqref{eq: average}. In both scenarios, all other particles are initially aligned with the $x$-axis, $\theta_{ij}(0) = 0$ for $ij \neq 00$. While (a) can be interpreted as a flock following a leader, (b) might describe a flock reacting to a sudden perturbation.

In the firm-leader scenario, the dynamical equation~\eqref{eq: average} is most easily written and solved using the matrix form ${\bm \theta}(t+1) = M_0 {\bm \theta}(t) +  {\bm \theta}(0)$, where the vector ${\bm \theta}(t)$ contains the values of orientations at all grid points at time $t$, $\theta_{ij}(t)$, and $M_0$ incorporates the  interactions. It has vanishing entries for  the feedback onto the leader's orientation, which is set by ${\bm \theta}(0)$, which has vanishing entries for all other particles. The solution is  ${\bm \theta}(t) = \sum_{i=0}^{t} M_0^i {\bm \theta}(0)$. In the lax-leader scenario, the dynamical equation is ${\bm \theta}(t+1) = M {\bm \theta}(t)$, and  $M$ incorporates the interactions between all the grid points, as described by Eq.~\eqref{eq: average}, including the feedback onto the leader. The solution is ${\bm \theta}(t) = M^t {\bm \theta}(0)$. Both solutions nicely demonstrate that due to the linearity of the dynamics, the transmission of the information obeys the principle of superposition: The impact onto  $\theta_{ij}(t)$ depends on the number of possible paths of length $t$ the signal may take from $(0,0)$ to $(i,j)$,
namely the summation induced by the matrix multiplication in $M_0^t$). And it decays with time and distance due to the conservation enforced by the repeated normalization via the prefactor $(1/5)^t$ in $M_0^t$.

\begin{figure}
    \centering
    \includegraphics[width=1.0\columnwidth]{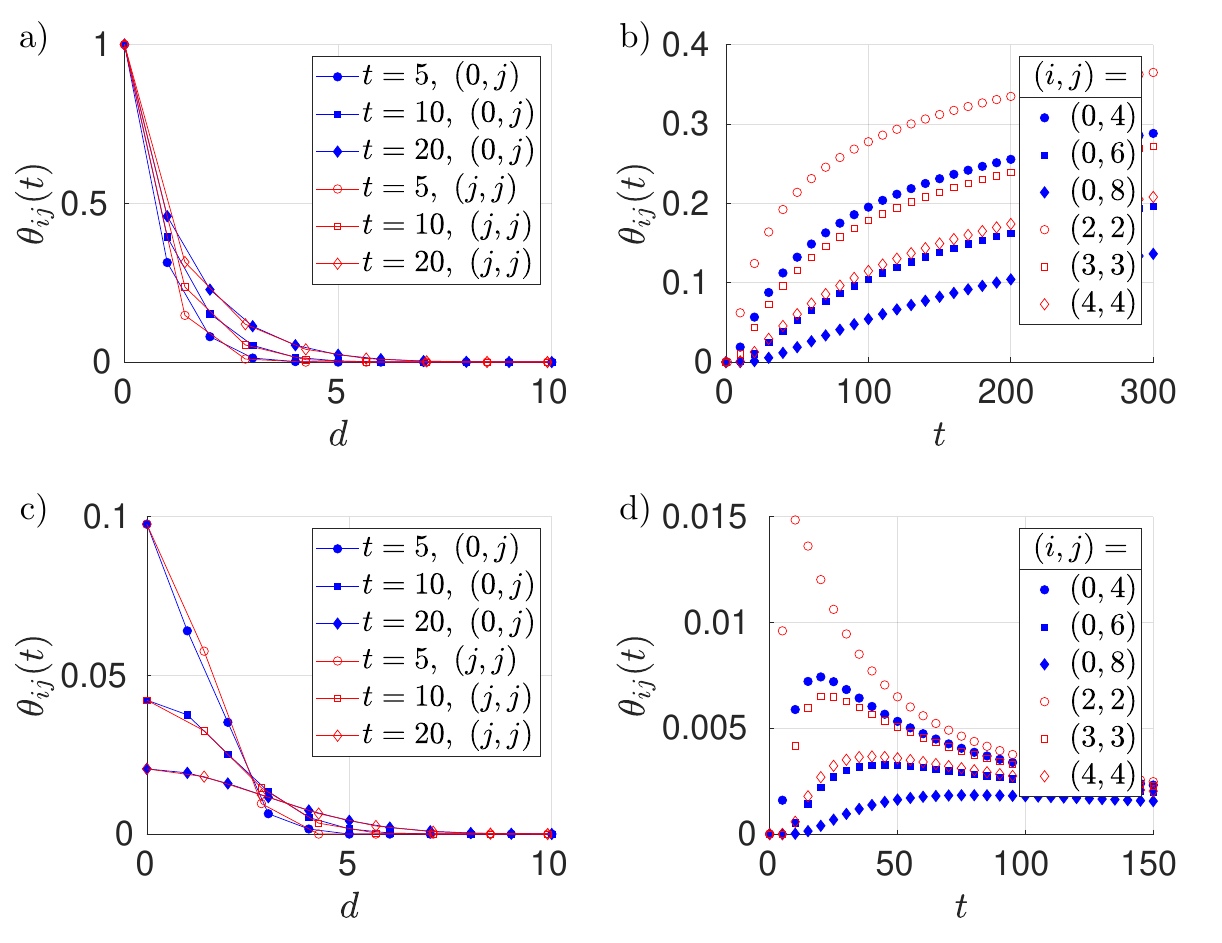}
	\caption{Information spreading in the firm (a-b) and lax (c-d) leader scenarios. a,c) The spatial spreading of the orientation $\theta_{ij}(t)$ at different times over the distance $d = \sqrt{i^2+j^2}$, transverse (filled blue markers) and diagonal (open red markers) relative to the leader. b,d) The time evolution of the orientation for different grid points.}
	\label{fig:signal spread}
\end{figure}

In Fig.~\ref{fig:signal spread} we depict the information spreading  in the \textcolor{black}{linearized} lattice VM for both scenarios. 
As expected, the information spreading quickly becomes isotropic, since discretizing the diffusion equation on a square lattice destroys the radial symmetry only for short paths and only affects the initial stage of the dynamics.  The spreading for the firm leader scenario, with a fixed source at the origin, eventually aligns all particles to the leader. The rate of this approach decreases with growing distance of the grid points from the leader and the saturation curves exhibit maximum slopes at intermediate times.

\subsection{Signal speed} 
\label{sec:SignalSpeedLM}

\begin{figure}
    \centering
    \includegraphics[width=1.0\columnwidth]{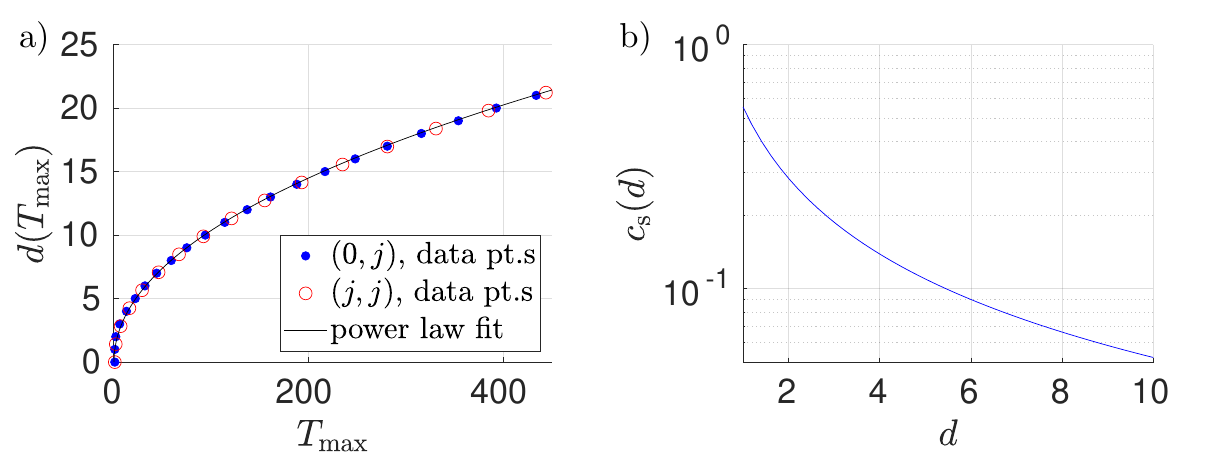}
	\caption{a) The dispersion relation for the \textcolor{black}{linearized} lattice VM transverse (blue filled circles) and diagonal (red open circles) relative to the leader. The solid line shows a fit $d(t)=at^m$ with exponent $m= 0.48$ and $a \approx 1.1$. b) The corresponding signal speed $c_{\rm s} = \dot{d}$ as function of the distance  $d$ to the leader.}
	\label{fig:dispersion relation1}
\end{figure}

In general, there is no unique definition of the speed of information spreading in the \textcolor{black}{linearized} lattice VM. The most obvious definition $v=l/\tau=1$ refers to the signal transfer between neighboring lattice points [cf Eqs.~\eqref{eq: average}--\eqref{eq: continuous}]. It provides the time $d/v$ after which a grid point at distance $d$ from the leader starts to receive the information. Yet, it is of limited use because the strength of the received information is negligible if the grid point is far from the leader and there are only a few paths for the signal between the leader and the grid point. For example, in the case of a single path the signal strength received at time $d/v$ is proportional to $(1/5)^d$. 
 
 A more informative definition is obtained from the time $T_\mathrm{max}(d)$ when the change of orientation induced by the leader  at distance $d$ becomes maximal. The rate of change of orientation of the grid points is measured by the time derivatives $\dot{\theta}_{ij}(t)$, which exhibit a clear maximum [cf.\ Figs.~\ref{fig:signal spread}b and d]. One may thus identify  $T_\mathrm{max}(d)$ with the time when $\dot{\theta}_{ij}(t)$ with $\sqrt{i^2 + j^2} = d$ is maximal. In Fig.~\ref{fig:dispersion relation1}a we show the resulting dispersion relation $d(T_\mathrm{max})$ obtained from evaluating the signal propagation on the horizontal and on the diagonal axis with respect to the leader in the firm leader scenario. As expected, the found information spreading is well described by the diffusion relation $d(T_\mathrm{max}) = \sqrt{4 D_{\rm eff} T_\mathrm{max}}$. However, the diffusion coefficient $D_{\rm eff} = a^2/4 \approx 0.3$, obtained by fitting the data, is much different than the diffusion coefficient $D = 0.2$, predicted from the limiting process leading to Eq.~\eqref{eq: parabolic}. In Fig.~\ref{fig:dispersion relation1}b, we show the corresponding signal speed $c_s = \dot{d}(t) \propto 1/\sqrt{t} \propto 1/d$. The results obtained for the lax leader scenario are qualitatively the same (data not shown).

As an aside, we note that, while evaluating the evolution of the maxima of $\dot{\theta}_i(t)$ is a reasonable approach for studying the signal spreading in the two flooding scenarios considered here, it is not suitable for more complex situations. A more universally applicable proxy for signal speed can be obtained by evaluating the connected acceleration correlations~\cite{cavagna2018physics}. For our specific setting with a single leader and aligned initial state, the two approaches lead to the same results. 

\begin{figure}
    \centering
    \includegraphics[width=1.0\columnwidth]{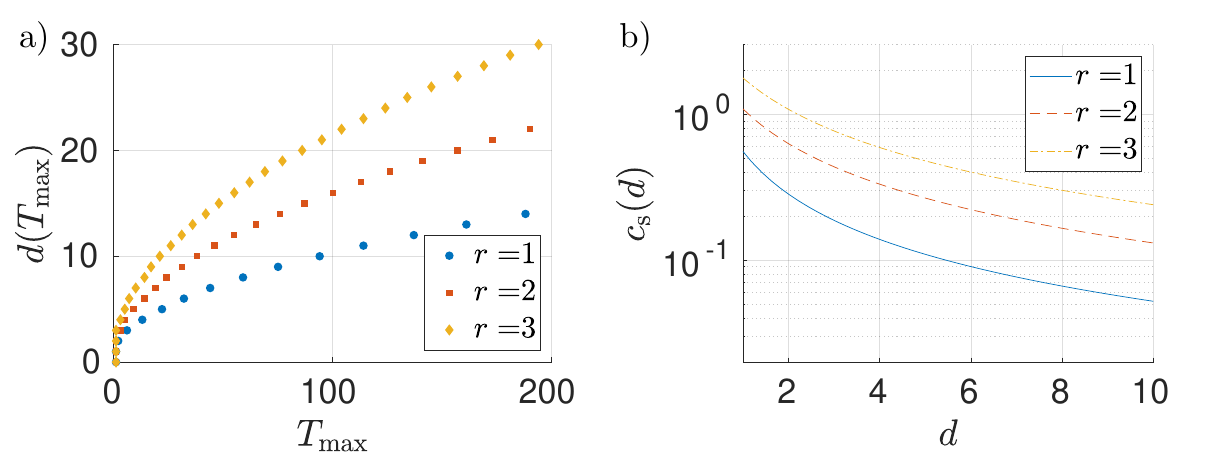}
	\caption{a)	The dispersion relation for a generalized lattice VM where each grid point interacts with all neighbors at distances up to $r$ lattice edges, for three values of $r$.
	b) The corresponding signal speeds $c_{\rm s}(d) = \dot{d}$, respectively.}	
	\label{fig:dispersion relation2}	
\end{figure}

To close this section, we investigate the information spreading in a direct generalization of the \textcolor{black}{linearized} lattice VM~\eqref{eq: average}, where the individual grid points do not interact only with their nearest neighbors, but with all grid points up to a distance of $r$ lattice edges from the leader. Consequently, each grid point interacts with its $N_\mathrm{int}=2r(r+1)$ nearest neighbors. 
The maximum speed of information propagation, $v$, is determined just by distances between particles at the circumference of the interaction zone and thus it increases linearly with $r$. On the other hand, the $r$-dependence of the speed $c_{\rm s}$, shown in Fig.~\ref{fig:dispersion relation2}, is sublinear as the signal maximum is `slowed down' by the particles inside the interaction radius. Interestingly, the curves for different $r$ values cannot be collapsed into a single master curve by multiplying each of them by a constant factor. Our analysis suggests that such a collapse is possible for long times only, with numerically obtained scaling factors 1.565 and 2.105 yielding the best asymptotic collapse of the curves for $r=1$ to those for $r=2$ and $r=3$, respectively. These factors are close to the factors 1.64 ($r=1\to2$) and 2.28 ($r=1\to3$) obtained from the diffusion limit~\eqref{eq: parabolic} of the individual lattice models as $\sqrt{D_r/D_1}$ with $D_r$ denoting the diffusion coefficient obtained for the individual values of $r$. Even though the diffusive scaling $d = \sqrt{4 D_r t}$, predicted from Eq.~\eqref{eq: parabolic}, does not describe the data perfectly (in particular the prefactor $4 D_r$ is wrong), we take this as an indication that the formula $d =\sqrt{4 D_{\rm eff} t}$ with $D_{\rm eff} \sim D_r$ is a reasonable qualitative model for the spreading of information over long time and large length scales. 

To sum up, in the \textcolor{black}{linearized} static spin model, the information spreads essentially diffusively. We next investigate how the situation changes when we allow particles to translate along their orientations.

\section{The motile case}
\label{sec:finite_velocity}

\subsection{The role of convection}

Compared to equilibrium systems, active matter breaks certain local symmetries such as momentum and energy conservation. It is not \emph{a priori} obvious, whether this fundamental difference will lead to  important effects on the information spreading and swarm behavior or is  largely irrelevant, in practice. Different from the 
lattice VM, the standard VM does not place the particles  onto a regular lattice, and if so, this order would not be maintained for long. Notice that this breaks two important symmetries, namely reciprocity and information conservation (see App.~\ref{sec:info_conserv}). While the disorder itself does not affect the diffusive information spreading, the evolution of the neighborhood relations for $v_0 > 0$ additionally allows for information convection. This situation is thus very similar to a moving heat source with the main difference that particles addressed by the leader tend to follow it, while heated \textcolor{black}{passive} particles generally do not induce a comparable flux.

Let us now derive a rough estimate for the particle speed $v_0$ at which convection becomes important. The maximum conduction speed is given by the speed with which the signal spreads due to the interactions, i.e., $\ell_\mathrm{int}/\Delta t$. Here $\Delta t = 1$ is the discrete update time in the VM and $\ell_\mathrm{int} = \sqrt{N_\mathrm{int}/(\pi\rho)}=\sqrt{N_\mathrm{int}/N}$ is the average interaction radius, assuming a more or less homogeneous density  $\rho = N/\pi$ after initiation inside a unit circle. The speed of convection is given by the relative speed of the individual particles on the order of $v_0$. Conduction and convection should thus compete when $v_0 \approx \sqrt{N_\mathrm{int}/N}$. 
Alternatively, as in Sec.~\ref{sec:SignalSpeedLM}, we could measure the speed of signal propagation by the ratio $d_i/T_i$, where is the time when the signal sent at time 0 causes a maximum change $\dot{\theta}_i(t)$ of orientation  at distance $d_i$, i.e., \textcolor{black}{ $\dot{\theta}_i(T_i)\equiv \max_t \dot{\theta}_i(t)$}. As we find below, the latter approach, which predicts a significantly lower conduction speed, is more appropriate to describe the data, yielding a correspondingly lower threshold velocity for the onset of convective transport. (For our choice of parameters, convection plays role already for velocities of about $v_0 = 0.01$ while $\sqrt{N_\mathrm{int}/N} \approx 0.16$.)

Besides inducing convection, motility further complicates the definition of a signal speed. Due to the relative motions of the particles there is no a priori choice of the distance $d_i$ travelled by a signal. For this reason, we analyzed the speed of information propagation using two different definitions of $d_i$. 
First, the (average) initial distance \textcolor{black}{$|\mathbf{r}_i(0)-\mathbf{r}_\mathrm{L}(0)|$} between the particles and the leader, which is the initial position of particle $i$ at time 0, also  encoded in the initial density $\rho$. Secondly, the distance \textcolor{black}{$|\mathbf{r}_i(T_i)-1/N\sum_j\mathbf{r}_\mathrm{j}(T_i)|$} between the particle $i$ and the position of the center (of mass) of the flock at the characteristic ``interaction  time'' $T_i$ for conductive transport. We have performed the analysis below for both these definitions of the distance and found no qualitative differences. Therefore, we only show  the results obtained for the former, in the following.

\subsection{Numerical procedure}

In the simulations, we place the leader always into the center of a unit circle. Positions of all  $N=1000$ other particles are picked randomly inside the circle. All particles interact with their $N_{\rm int}=24$ nearest neighbors, corresponding to $r=3$ in Fig.~\ref{fig:dispersion relation2}. Small density fluctuations in the initial condition are found to induce strong noise in the measured functions $\theta_i(t)$ and $\dot{\theta}_i(t)$. To be able to determine the overall trend from these measurements, we averaged the resulting curves over $N_{\rm runs}$ runs with different initial conditions. Besides, we employed two different smoothening procedures:

The average $\langle \bullet \rangle_{nn}$ is calculated as follows. First, we collect the data $\{d_i,\theta_i(t),\dot{\theta}_i(t)\}_{i=1,...,N}$ from $N_\mathrm{runs} = 100$ runs of the simulation.  Then, we sort the data according to the distance $d_i$ to the origin at time 0. And finally, we calculate the smoothened variables $\langle d_i \rangle_{nn}$, $\langle\theta_i(t)\rangle_{nn}$, and $\langle\dot{\theta}_i(t)\rangle_{nn}$ by averaging $d_i$, $\theta_i(t)$, and $\dot{\theta}_i(t)$ over $N_\mathrm{av}$ neighbors of the particle $i$, i.e., over the particles $j$ with $N_\mathrm{av}$ smallest distances $|d_i - d_j|$. Since the dispersion relation is a strictly monotonous function of time, one may alternatively perform the averaging with respect to nearest neighbors in time $T_i$, according to when the maximum signal has arrived at particle $i$. In other words one can average over the particles $j$ with $N_\mathrm{av}$ smallest distances $|T_i - T_j|$. We have tested that both averaging procedures lead to qualitatively the same results. In the following, we only show those obtained using the averaging $\langle \bullet \rangle_{nn}$ over $N_\mathrm{av}$ spatially nearest neighbors.

\begin{figure}
    \centering
    \includegraphics[width=1.05\columnwidth]{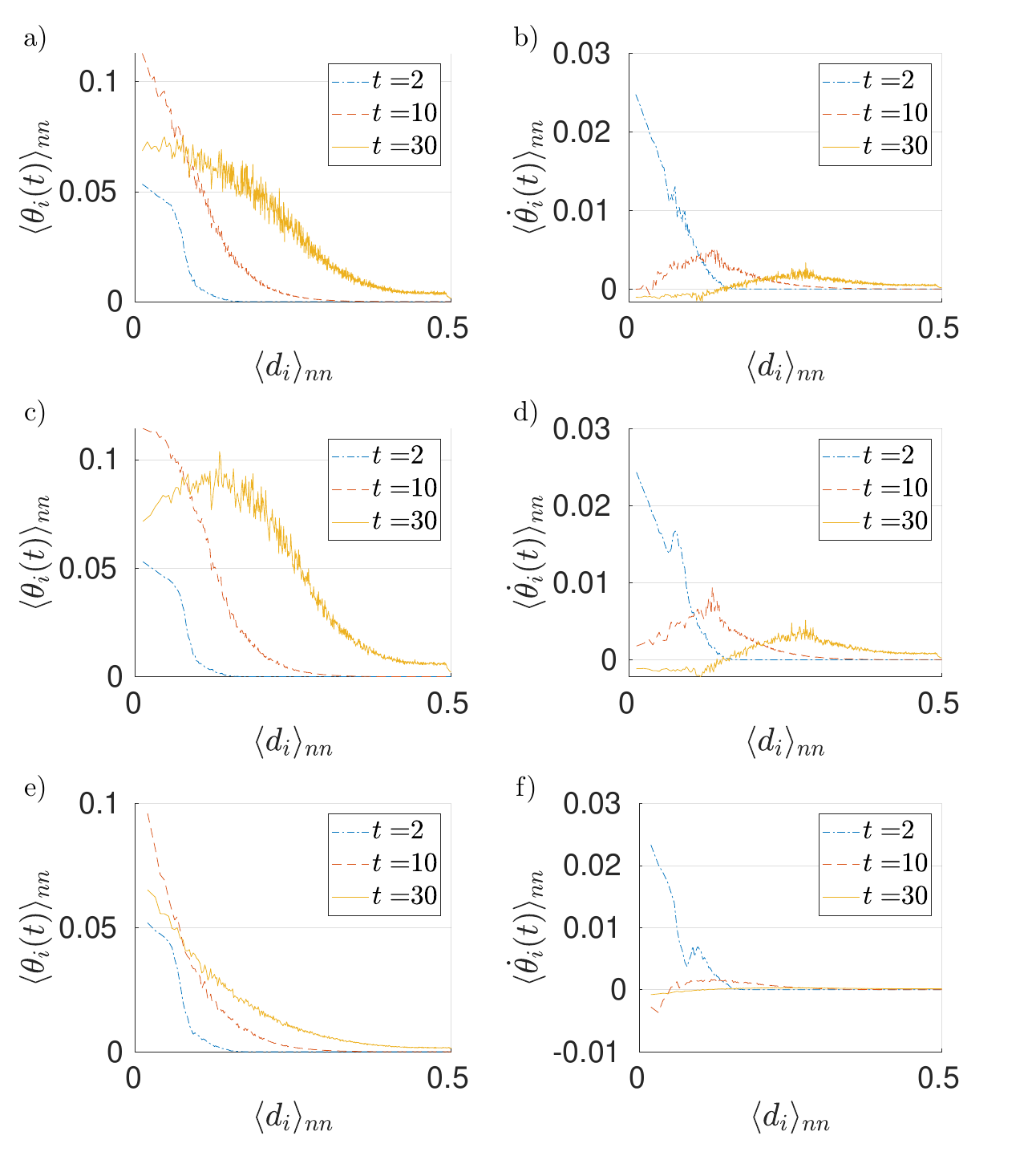}
	\caption{Firm leader scenario. The average orientation, $\langle\theta_i(t)\rangle_{nn}$ and change in orientation $\langle\dot{\theta}_i(t)\rangle_{nn}$, as functions of the average distance $\langle d_i \rangle_{nn}$ at three different times: for the whole swarm (a-b), along the leader direction (c-d), and along the negative leader direction (e-f). Parameters used: $N=1000, v_0=0.01, N_{\rm int}=24, \varphi=\pi/4, N_\mathrm{runs}=100, N_\mathrm{av}=N_\mathrm{runs}$. }
	\label{fig:VM spatial}	
\end{figure}

\subsection{Firm leader scenario}

We now consider the firm leader scenario of Sec.~\ref{Sec: LM Numerical}, where the leader's orientation is fixed to $\varphi$ at all times and all other particles are initially aligned with the perpendicular $x$-axis and subsequently obey the dynamical equations~\eqref{eq:vtdisc}-\eqref{eq:rtdisc}. Note that this condition implies that reciprocity between the leader and the flock is maximally broken.

In Figs.~\ref{fig:VM spatial}a-b we show the resulting averaged orientations, $\langle\theta_i(t)\rangle_{nn}$, and the averaged changes in the orientation, $\langle\dot{\theta}_i(t)\rangle_{nn}$, as functions of the averaged distance $\langle d_i \rangle_{nn}$.  To investigate the directional dependence of the information spreading, we distinguish between two directions of signal propagation. As the leaders orientation point's into the positive half-plane, we identify the particles with positive $y$-coordinates at time $0$ as lying in the `positive direction' with respect to the leader. The remaining particles are lying in the `negative direction'. The results for $\langle\theta_i(t)\rangle_{nn}$ and $\langle\dot{\theta}_i(t)\rangle_{nn}$ for the positive and negative direction are given in Figs.~\ref{fig:VM spatial}c-d and e-f, respectively. As the leader carries the source of information with it, particles lying in positive direction show a significantly larger change of orientation than those in the negative direction. 
\textcolor{black}{Furthermore, the leader affects nearby particles more than more distant ones. This leads to correspondingly stronger average direction changes $\langle \dot{\theta}_i(t)\rangle_{nn}$ in its vicinity. Consequently, upon traversing the flock, the leader seems to drag around a cloud of ``followers'’. However, since the interaction rule allows only imperfect alignments, particles begin to realign with the less informed surroundings after the leader has left their neighborhood. This is depicted by the moving maxima of $\langle \theta_i(t)\rangle_{nn}$ and $\langle \dot{\theta}_i(t)\rangle_{nn}$ in panels c) and d). In the negative direction, where the information propagates by pure conduction, no such structure is visible.}
The response of the swarm as a whole is dominated by the dynamics in positive direction. 
Repeating the described analysis for $v_0 = 0$, we found the same behavior as for the \textcolor{black}{linearized} lattice VM in Sec.~\ref{Sec: Lattice model}.

\begin{figure}
    \centering
    \includegraphics[width=1.05\columnwidth]{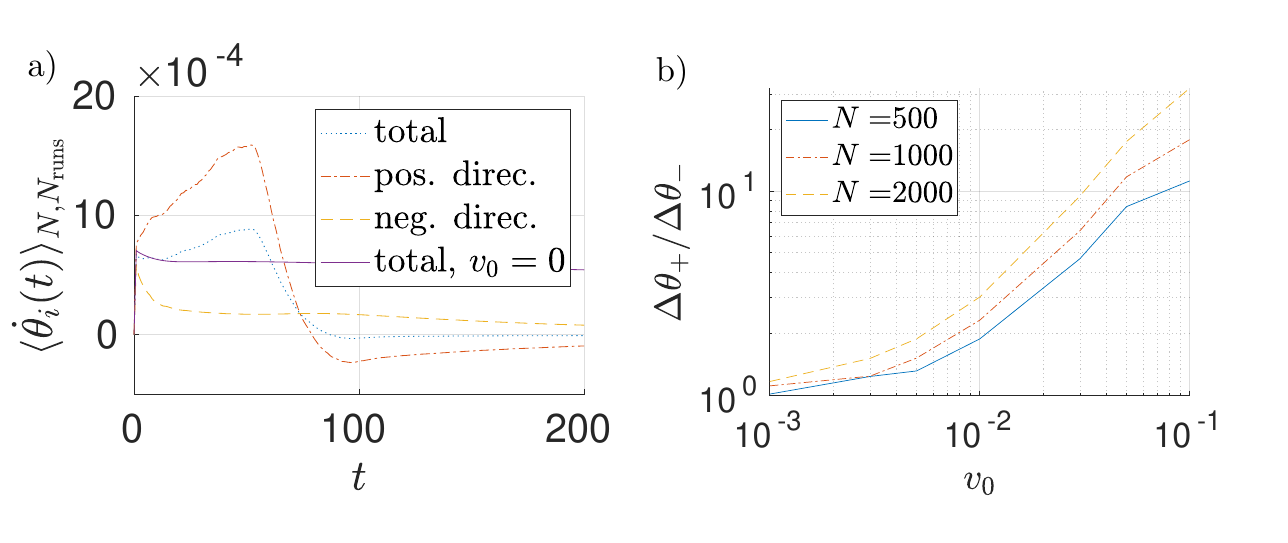}
    \caption{Firm leader scenario. a) Time evolutions of the change of direction averaged over the total system and the positive and negative directions for $v_0=0.01$ and $v_0=0$, respectively. b) The corresponding ratio~\eqref{eq:Deltapm}  as function of the particle speed. Other parameters  as in Fig.~\ref{fig:VM spatial}.}
    \label{fig:VM time}	
\end{figure}

In Fig.~\ref{fig:VM time}a, we show the time evolution of the change of orientation $\langle \dot{\theta}_i(t)\rangle_{N,N_\mathrm{runs}}$ averaged over all particles in the chosen particle set (total system, positive direction, and negative direction) and all simulations for $v_0>0$ and $v_0=0$. 
For $v_0>0$, the average signal strength for the total system and particularly in the positive direction continuously increases until the leader approaches the edge of the system. This can be understood as follows. The change of orientation of the individual particles is largest when their orientation differs most \textcolor{black}{from} the average orientation of their neighbors. A moving leader constantly meets on its way in the positive direction particles (almost) aligned with the $x$-axis, leading to a steady increase of the corresponding signal. 
\textcolor{black}{Subsequently, at times $t \gtrsim 50$ when the leader has left the flock,  $\langle \dot{\theta}_i(t)\rangle_{N,N_\mathrm{runs}}$ rapidly decreases and eventually it also changes sign. Since the leader mainly affects nearby particles, more distant particles are much less aligned with its orientation when it leaves. Therefore, particles that aligned with the leader during its passage through the flock begin to realign with the less affected particles after the leader has left.
} 
In the negative direction, the signal strength monotonously decreases, similarly as for $v_0=0$ and the \textcolor{black}{linearized} lattice VM. 

To quantify the asymmetry between the positive and negative direction, we integrate the positive areas 
\begin{equation}
\Delta \theta_\pm = \int_{t_0}^{\infty} \langle \dot{\theta}_i(t) \rangle^{\pm}_{N,N_\mathrm{runs}} \Theta(\langle \dot{\theta}_i(t) \rangle^{\pm}_{N,N_\mathrm{runs}}),
\label{eq:Deltapm}
\end{equation}
beneath the corresponding curves in  Fig.~\ref{fig:VM time}a.
Here $+$ (-) corresponds to the positive (negative) direction and $\Theta(\bullet)$ denotes the unit step function. The ratio $\Delta \theta_+/\Delta \theta_-$ is shown in Fig.~\ref{fig:VM time}b.
As expected, it monotonously increases with the particle speed $v_0$ and particle density $N/\pi$.

\begin{figure}
    \centering
    \hspace*{-0.6cm}
    \includegraphics[width=1.2\columnwidth]{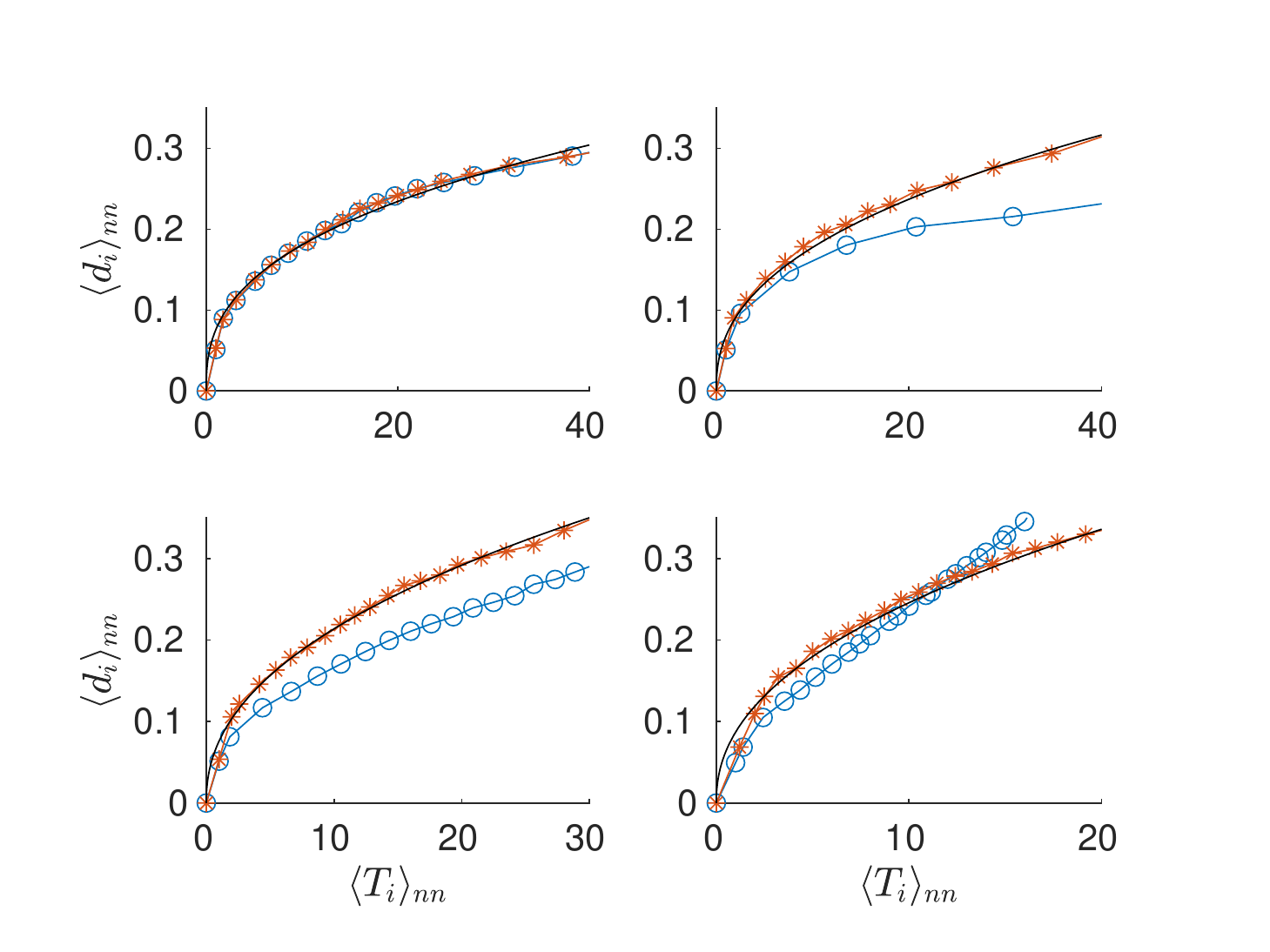}
    \caption{Dispersion relation for the firm-leader scenario for positive (blue circles) and negative (red stars) directions and speeds $v_0= 0, 0.001, 0.01, 0.03$ increasing from the upper left to the bottom right panel. The black solid lines are fits of $a \left<T_i\right>_{nn}^m$ to the data for the positive direction with $m=0.376, 0.417,  0.449, 0.451$ corresponding to the individual speeds. The slope of the data for speed $v_0 = 0.03$ for the positive direction is $\approx 0.018$ while the corresponding speed of the leader projected to the positive direction is $v_0 \sin \varphi \approx 0.021$.  Parameters as in Fig.~\ref{fig:VM spatial}, 
    except for $N_\mathrm{Av}=10N_\mathrm{runs}$.}
    \label{fig:VM_DR_compare_v_0}
\end{figure}

The main result of this section are the dispersion relations for four different velocities shown in Fig.~\ref{fig:VM_DR_compare_v_0}. Regardless of $v_0$, the information initially spreads conductively, hence similarly as in the \textcolor{black}{linearized} lattice VM, as diffusion beats convection over short times and distances.  With increasing velocity $v_0>0$, the spreading  in the positive direction becomes gradually more convective, at late times. The slope of the dispersion relation converges to the velocity of the leader projected to the positive direction, $v_0 \sin \varphi$.
In the negative direction, the spreading stays conductive regardless of $v_0$. Even though the particles in the negative direction are less affected by the turning event induced by the leader, the dispersion relation shows that the information reaches them faster than those in the positive direction. This counter-intuitive effect is a consequence of the employed definition of $T_i$: the dispersion relation follows from determining times maximizing $\dot{\theta}_i(t)$. As the leader moves away from the particles behind it, the rates of change  $\dot{\theta}_i(t)$ of their orientations peak sooner than those in the positive direction. This somewhat counter-intuitive behavior is reminiscent of observations of faster speeds for smaller pulses~\cite{joseph1989heat} or propagation of second sound against the heat flow~\cite{coleman1983nonequilibrium}. 

\subsection{Lax leader scenario} \label{app: alternative}

\begin{figure}
    \centering
    \includegraphics[width=1.05\columnwidth]{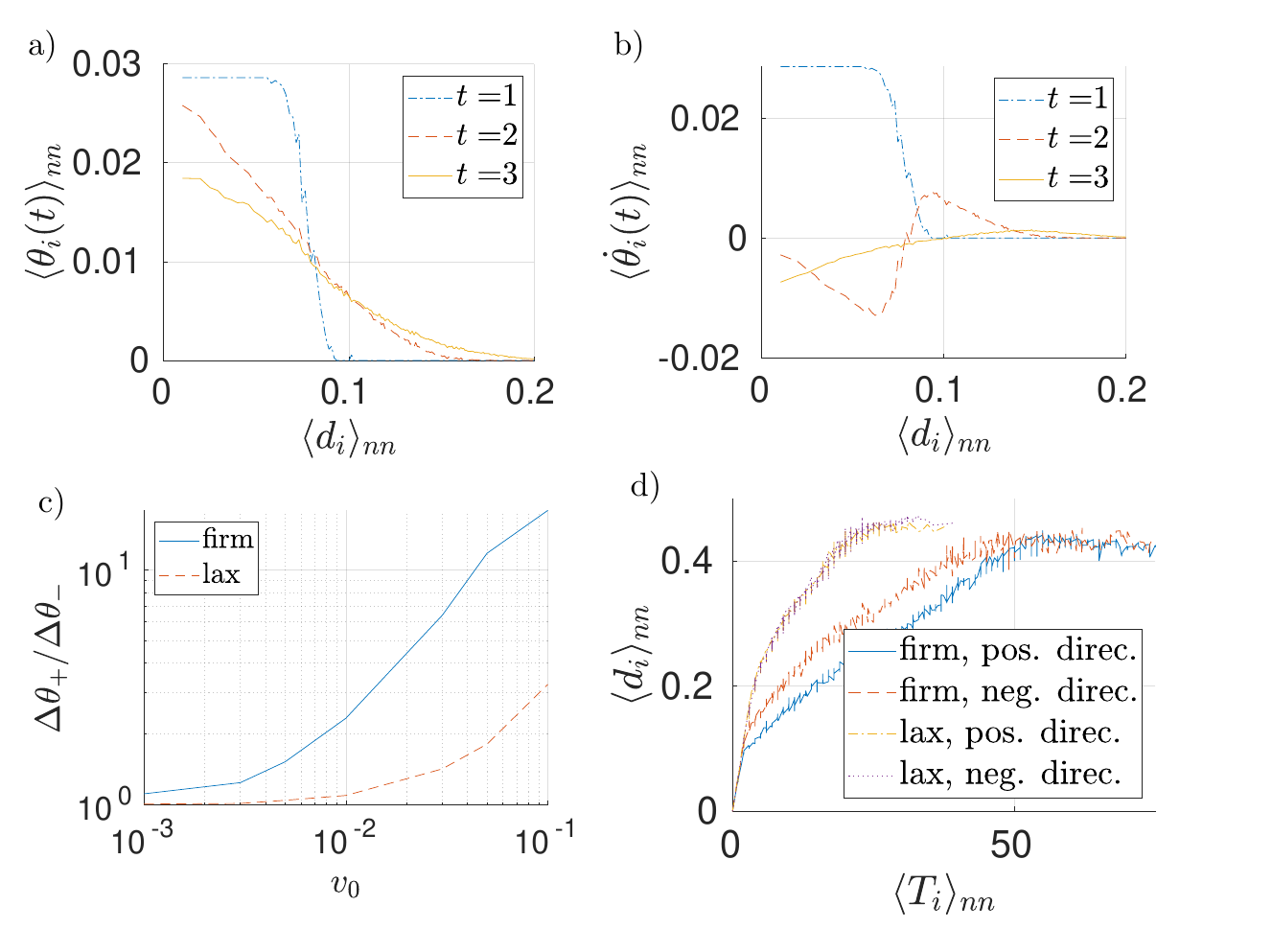}
	\caption{Lax leader scenario. a-b) The orientation, $\langle\theta_i(t)\rangle_{nn}$, and the averaged change in the orientation, $\langle\dot{\theta}_i(t)\rangle_{nn}$, as functions of the averaged distance $\langle d_i \rangle_{nn}$ for three different times averaged over the overall system. c) The ratio~\eqref{eq:Deltapm} of the integrated positive changes in the orientation in the positive and negative direction as functions of the particle speed for the firm and lax leader scenario. (d) The dispersion relation for the positive and negative directions for the firm and lax leader scenario. We used the same parameters as in Fig.~\ref{fig:VM spatial}.}
	\label{fig:VM spatial_pert2}	
\end{figure}


The information spreading is somewhat different in the lax leader scenario. For the same initial condition as in the preceding section, the leader now adapts dynamically according to equations~\eqref{eq:vtdisc} and \eqref{eq:rtdisc} to its neighbors for $t>0$. It thereby virtually \textcolor{black}{loses} the information passed on to them.  The interaction with the neighbors is thus more reciprocal than for the unwavering firm leader, yet not entirely so, since the topological notion of next neighbors is not necessarily fully reciprocal (See Sec.~\ref{sec:info_conserv}). 

In Fig.~\ref{fig:VM spatial_pert2}a-b, we depict again the average orientation $\langle\theta_i(t)\rangle_{nn}$ and the averaged change in the orientation $\langle\dot{\theta}_i(t)\rangle_{nn}$ for the lax leader scenario. The parameters are the same as for the firm leader scenario in Fig.~\ref{fig:VM spatial}. Comparing the results for the two scenarios, we find the following differences:  
\begin{enumerate*}[label=(\roman*)]
    \item the amplitudes of both $\langle\theta_i(t)\rangle_{nn}$ and $\langle\dot{\theta}_i(t)\rangle_{nn}$ are much smaller and the time derivative in the averaged angle converges to zero much faster, since the leader realigns with the rest of the flock in the lax leader scenario;
    \item The time derivative $\langle\dot{\theta}_i(t)\rangle_{nn}$ exhibits an excursion to negative values at small distances to the leader, due to the feedback from the flock, which requires a realignment of the leader and its neighborhood with the ``winning majority'' of other particles in the flock. At larger distances, the derivative returns to positive values, as expected for a moderate realignment of the merely slightly disturbed more distant particles.
    \item There is again a directional dependence of the response, as in the firm-leader scenario. However, it is now much weaker, due to the mutual information exchange.
\end{enumerate*}

In fact, for the parameters used in the figure, it is not worthwhile to show the corresponding spatial distributions, as they would be hardly descernible from those for the total system in Fig.~\ref{fig:VM spatial_pert2}. The directional dependence of the information spreading in the lax leader scenario only becomes noteworthy for substantially larger speeds $v_0$, as demonstrated in Fig.~\ref{fig:VM spatial_pert2}c. There we compare the ratio $\Delta \theta_+/\Delta \theta_-$ of responses~\eqref{eq:Deltapm} integrated in the positive and negative directions between the lax and firm leader scenarios. 
Because of weaker total signal strength in the lax leader scenario, the changes of orientation  $\dot{\theta}_i(t)$ of the individual particles peak sooner, leading to steeper (but directionally barely distinguishable) dispersion relations, cf.\ Fig.~\ref{fig:VM spatial_pert2}d. 

\section{Conclusion} \label{sec:Conclusion}

We studied transport of information about orientation of a leader in the Vicsek model (VM) with topological interactions.
The two main mechanisms for propagation of information are conduction and convection. We have shown that the conductive aspect in the VM can well be understood using a simplified, exactly solvable variant of the model, where the individual particles are fixed at grid points of a regular lattice. This static spin lattice model allows to draw an analogy with heat transfer, which ceases to hold in the full dynamic VM. \textcolor{black}{Nonlinearity and heterogeneity then break the underlying symmetries such as reciprocity and spin conservation.} Yet, this has no major practical consequences, by itself. The visible changes between the dynamic model and the spin lattice are entirely dominated by the convective dynamics.

We considered two scenarios of information spreading from a single misaligned leader particle. While the diffusive/conductive spreading prevails over short times and distances, the spreading over longer times and distances gradually acquires a convective contribution, as the particle speed increases. We quantified this intuitive conclusion by measuring the dispersion relation. It was formulated for the time scale at which the signal induces its largest change in orientation of the particles at a given distance. The analysis revealed a strong directional dependence of the information transfer for the firm-leader scenario, in which the reciprocity of the mutual information exchange is maximally violated. A significant effect of convective information spreading is observed only in the direction of the leader motion. In the wake zone behind the leader, the spreading remains diffusive, regardless of the speed.

While measuring the dispersion relation for zero speed of the particles is a relatively straightforward task, the definition of the distance over which the signal has propagated becomes ambiguous for the motile swarm. Nevertheless, we found that different length definitions lead to qualitatively close results.

Besides this ambiguity in the definition of the dispersion relation, which might deserve further analysis, our findings raise several questions. First, while some preliminary runs seemed to confirm the expectation that the inclusion of noise in the VM would yield qualitatively similar results, one could wish to study this issue more extensively, in particular with regard to the stability of the flocking transition; i.e., under which conditions can a leader move induce an ordering transition or the breakup of a flock?
Further, many natural interaction networks are more heterogeneous than our flocks, containing, e.g., certain hierarchical structures~\cite{chase1974models,nagy2010hierarchical} or distance/density weighted interactions~\cite{vicsek1995novel}. \textcolor{black}{Moreover, it might be interesting to consider a more realistic modification of the standard VM where the orientation of a particle under consideration would have a stronger weighting than the average orientation of its neighbors. This would yield a more persistent motion and might impact the information propagation.} We took first steps in this direction in a follow-up study to the present work~\cite{geiss2021Vicsek}.  There, we investigate information spreading in a 2d VM with time-delayed metric interactions~\cite{Holubec2021}, and also address the pertinence of the notion of linear response~\cite{czirok1997spontaneously,chate2008collective,pearce2016linear} and its relation to information propagation. Next, it might be interesting to connect information spreading in active matter with corresponding results in other research fields, such as network theory or epidemiology. Especially in the latter, the effects of network heterogeneity on the spreading of diseases is a widely studied aspect~\cite{olinky2004unexpected,meyers2005network,grossmann2021ode,levis2020flocking}. Finally, it would seem interesting to pursue the question, how the interaction rules in the VM can be optimised to facilitate information transfer.

\section*{Acknowledgments}
We acknowledge funding through
a DFG-GACR cooperation by the Deutsche Forschungsgemeinschaft (DFG Project No 432421051) and by
the Czech Science Foundation (GACR Project No 20-02955J). VH was supported by the Humboldt Foundation. DG acknowledges funding by International
Max Planck Research Schools (IMPRS) as well as by Deutscher Akademischer Austauschdienst (DAAD).

\appendix

\section{Breaking of information conservation in the \textcolor{black}{linearized} VM}
\label{sec:info_conserv}

As discussed in Sec.~\ref{Sec: Lattice model}, the interaction rule in the \textcolor{black}{linearized} lattice VM,
\begin{equation}
    \theta_{ij}(t+1) = \frac{1}{N_{ij}(t)} \sum_{\langle(ij),(kl)\rangle} \theta_{kl}(t),
\end{equation}
where the sum goes over all neighbors $(kl)$ of $(ij)$ including $(ij)$ itself, and $N_{ij}=\sum_{\langle(ij),(kl)\rangle}$ denotes the number of neighbors, conserves the total information content
\begin{equation}
\theta_\mathrm{tot}(t) \equiv \sum_{ij} \theta_{ij}(t).
\label{eq: conservation}
\end{equation}
While the \textcolor{black}{linearized} Vicsek interaction rule yields reciprocal interactions if the interaction network is regular, it can render non-reciprocal inter-particle interactions for irregular interaction networks, e.g., if the particle density of the system is inhomogeneous in space. The conservation condition $\dot{\theta}_\mathrm{tot}(t) = 0$ will then be broken. For a closed system, the reverse also holds: if the conservation is broken, this indicates the presence of some non-reciprocal interactions. As an illustrative example, consider the following closed interaction network consisting of three particles. Particle 1 and 3 interact solely with particle 2, which interacts with both 1 and 3. Assuming the initial condition $\theta_1(0)=1$ and $\theta_{2,3}(0)=0$, then $\theta_1(1)=1/2$, $\theta_2(1)=1/3$, $\theta_3(1)=0$. We thus see that $\theta_\mathrm{tot}(0) = 1 > \theta_\mathrm{tot}(1) = 5/6$. If we instead consider the initial condition $\theta_2(0)=1$ and $\theta_{1,3}(0)=0$, we find $\theta_\mathrm{tot}(0) = 1 < \theta_\mathrm{tot}(1) = 3/2$. These examples manifest a more general finding that $\theta_\mathrm{tot}(t)$ decreases when the information flows from a less to a more connected region, and vice versa. 
In this case, reciprocity is broken since the normalization $N_{ij}$ of neighboring particles varies. \textcolor{black}{Beyond the linear regime, the situation is more complicated as the normalization $N_{ij}$ depends on the angular variables.}

Similarly,  also for topological interactions, the information content is not conserved. While each particle interacts with exactly the same number of neighbors (i.e. $N_{ij}=\mathrm{const}$.), density gradients may induce unilateral interactions. As an example, consider a closed system of four particles with topological interactions with two nearest neighbors. Let the particles 1, 2 and 3 reciprocally communicate with each other, while the distant particle 4 adjust its direction to that of particle 2 and 3 without influencing them. Assuming the initial condition $\theta_2(0)=1$ and $\theta_{1,3,4}(0)=0$, we find $\theta_{1,\dots,4}(1)=1/3$ and thus $\theta_\mathrm{tot}(0) = 1 < \theta_\mathrm{tot}(1) = 4/3$.

\bibliographystyle{abbrv}
\bibliography{bibfile}

\end{document}